\documentclass[aps,prb,showpacs,twocolumn,superscriptaddress]{revtex4}

\usepackage{graphicx}

\usepackage{dcolumn}
\usepackage{bm}
\usepackage{latexsym}
\usepackage{amssymb}
\usepackage{amsfonts}
\usepackage{amsmath}
\usepackage{times}
\usepackage[colorlinks,bookmarks=false,citecolor=blue,linkcolor=red,urlcolor=blue]{hyperref}
\usepackage{color}

\begin{document}

\title{Competition between two- and three-sublattice ordering for $S=1$ spins on the square lattice}

\author{Tam\'as A. T\'oth}
\affiliation{Department of Physics, University of Toronto, Toronto, Ontario M5S 1A7, Canada}
\affiliation{Institut de th\'eorie des ph\'enom\`enes physiques,
Ecole Polytechnique F\'ed\'erale de Lausanne, CH-1015 Lausanne, Switzerland}
\affiliation{Research Institute for Solid State Physics and Optics, H-1525 Budapest, P.O. Box 49, Hungary}
\author{Andreas M. L\"auchli}
\affiliation{Institut f\"ur Theoretische Physik, Universit\"at Innsbruck, A-6020 Innsbruck, Austria}
\affiliation{Max Planck Institut f\"ur Physik komplexer Systeme, D-01187 Dresden, Germany}
\author{Fr\'ed\'eric Mila}
\affiliation{Institut de th\'eorie des ph\'enom\`enes physiques,
Ecole Polytechnique F\'ed\'erale de Lausanne, CH-1015 Lausanne, Switzerland}
\author{Karlo Penc}
\affiliation{Research Institute for Solid State Physics and Optics, H-1525 Budapest, P.O. Box 49, Hungary}

\date{\today}

\begin{abstract}
We provide strong evidence that the $S=1$ bilinear-biquadratic Heisenberg model with nearest-neighbor interactions on the square lattice possesses an extended three-sublattice phase induced by quantum fluctuations for sufficiently large biquadratic interactions, in spite of the bipartite nature of the lattice. The argumentation relies on exact diagonalizations of finite clusters and on a semiclassical
treatment of quantum fluctuations within linear flavor-wave theory.
In zero field, this three-sublattice phase is purely quadrupolar, and upon increasing
the field it replaces most of the plateau at 1/2 that is predicted by the classical theory.
\end{abstract}

\pacs{
75.10.Kt,  
75.30.Kz,  
75.10.Jm  
}

\maketitle
On bipartite lattices, classical magnetic models with nearest-neighbor interactions usually display
very simple types of order. Indeed, once the configuration that minimizes the
energy of a pair of sites
is known, the total energy can be minimized by extending the solution to the two
sublattices of the bipartite lattice. If both sites of the pair are in the same
state, the resulting configuration is uniform; otherwise it corresponds to a two-sublattice
state. This is true for Ising, {\it XY}, and Heisenberg models, with a uniform ferromagnetic ground state
for negative coupling and a two-sublattice N\'eel ground state for positive coupling. This reasoning
can be extended to quantum models by considering product wavefunctions (see below), and if quantum fluctuations
around this ``classical'' solution do not destroy the order, the ground state can be expected to be uniform or N\'eel-like. This is indeed known to be the case for the Heisenberg model for any spin. One might naively assume that this simple argument also applies to the generic spin-1 SU(2)-invariant model, the
bilinear-biquadratic Hamiltonian
\begin{equation}
  \mathcal{H} =  \sum_{\langle i,j\rangle} J_1 {\bf S}_i \cdot {\bf S}_j
  + J_2 \left({\bf S}_i \cdot {\bf S}_j\right)^2\;,    \label{eq:Hbb}
\end{equation}
since although the interaction includes a biquadratic term, it is still limited to nearest neighbors.

However, this picture is challenged by two observations. First of all, it is well established by now that
in the regime $0<J_1\le J_2$, the correlations of the 1D case are algebraic with period 3.~\cite{FathPeriod3,ItoiKato,AML_BLBQ} This indicates
that weakly coupled chains will be driven to a long-range ordered state that respects this periodicity, and this might also occur in the isotropic 2D case.
Second, our recent study of the SU(3) Heisenberg model on the square lattice~\cite{toth2010} has shown that three-flavor stripe order is stabilized, and this model
is equivalent to the model of Eq.~(\ref{eq:Hbb}) with $J_1=J_2$ in zero field. This raises a
natural question: Is the three-sublattice order a consequence of the enhanced SU(3) symmetry, or does
the model of Eq.~(\ref{eq:Hbb}) actually possess a phase with this symmetry?

In this Rapid Communication, we show that quantum fluctuations actually stabilize a three-sublattice phase in an unexpectedly
large portion of the phase diagram, including a parameter range in zero field where chains are gapped
with incommensurate short-range correlations,~\cite{Schollwock} and a finite-field region where
classical spins form a two-sublattice magnetization plateau at 1/2.

In order to set the stage, let us begin with a review of the classical (Hartree) solution, a variational approach based
on a product wave function of the form
\begin{equation}
 | \Psi \rangle =   \prod_{i=1}^N | \psi_i\rangle \;,
  \label{eq:Psivari}
\end{equation}
where $N$ is the number of lattice sites. The minimization of the energy with respect to the local wave functions $|\psi_i\rangle$ was achieved
by Papanicolaou some time ago.~\cite{papanicolaou88} Given the importance of the results for the discussion of
quantum effects, let us present the solution in a slightly more compact way than Ref.~\onlinecite{papanicolaou88} does.
In order to discuss general spin-1 states, it is convenient to introduce the time-reversal invariant basis
\begin{align}
|x\rangle &= i \frac{|1\rangle-|\bar 1 \rangle}{\sqrt{2}}\;,&
|y\rangle &= \frac{|1\rangle+|\bar 1 \rangle}{\sqrt{2}}\;,&
|z\rangle &= -i |0\rangle\;.
\end{align}
A general state $ |{\bf d}\rangle=\sum_\alpha d_\alpha |\alpha\rangle$ can be described by a complex
vector ${\bf d} = {\bf u} + i {\bf v}$ \cite{ivanov03}. Without loss of generality, one can choose $\bf u$ and $\bf v$
in such a way that ${\bf u} \cdot {\bf v}=0$, and the normalization imposes $u^2 + v^2=1$. If $u=v$, the
state is purely magnetic and $\langle {\bf S} \rangle=2 \ {\bf u} \times {\bf v}$. If $u=0$ or $v=0$, the
state is purely quadrupolar with a director along the nonzero component $\bf u$ or $\bf v$: It is obtained
by an SU(2) rotation from $|0\rangle$ and it displays spin fluctuations in a plane perpendicular to its director. Following the
general strategy for bipartite lattices, we first minimize the variational energy of a pair of sites: It is given by
\begin{equation}
E=J_1 |{\bf d}_i \cdot {\bf d}_j^*|^2 + (J_2-J_1) |{\bf d}_i \cdot {\bf d}_j|^2+J_2\;.
\end{equation}
This energy is minimum for configurations with $|{\bf d}_i \cdot {\bf d}_j^*|^2$ and
$|{\bf d}_i \cdot {\bf d}_j|^2$ equal to 0 or 1 (it turns out that the extreme values can be attained
simultaneously) depending on whether $J_1$ and $J_2-J_1$ are positive or negative. This
leads to the following solutions:

i) $J_1>0$, $J_2-J_1<0$: ${\bf d}_j={\bf d}_i^*$ to maximize $|{\bf d}_i \cdot {\bf d}_j|^2$, and $u_i=v_i$
to satisfy ${\bf d}_i \cdot {\bf d}_j^*=u_i^2-v_i^2=0$. The states are purely magnetic and
the spins are antiparallel. Note that this case includes the simple antiferromagnetic Heisenberg coupling.

ii) $J_1<0$, $J_2-J_1<0$: ${\bf d}_j={\bf d}_i^*$ to maximize $|{\bf d}_i \cdot {\bf d}_j|^2$, and $u_i=0$ or $v_i=0$ to maximize $|{\bf d}_i \cdot {\bf d}_j^*|^2=(u_i^2-v_i^2)^2$. The states
are purely quadrupolar with parallel directors.

iii) $J_1<0$, $J_2-J_1>0$: ${\bf d}_j={\bf d}_i$ to maximize $|\bf d_i \cdot \bf d_j^*|^2$, and $u_i=v_i$ to
satisfy ${\bf d}_i \cdot {\bf d}_j=u_i^2-v_i^2=0$. The states are purely magnetic with parallel spins. This case includes the simple ferromagnetic coupling.

iv) $J_1>0$, $J_2-J_1>0$: ${\bf d}_i \cdot {\bf d}_j={\bf d}_i \cdot {\bf d}_j^*=0$ implies ${\bf u}_i \cdot {\bf u}_j=
{\bf u}_i \cdot {\bf v}_j={\bf v}_i \cdot {\bf u}_j={\bf v}_i \cdot {\bf v}_j=0$. Since ${\bf u}_i \cdot {\bf v}_i=
{\bf u}_j \cdot {\bf v}_j=0$ by definition, the four vectors ${\bf u}_i$, ${\bf v}_i$, ${\bf u}_j$, and ${\bf v}_j$ must
be pairwise orthogonal. This implies that at least one of them vanishes; i.e., one state is a pure
quadrupole. The other state is only constrained by the condition that its $\bf u$ and $\bf v$ vectors must be perpendicular
to the director of the first state. It can be purely magnetic with a moment along this director, purely quadrupolar with a director perpendicular to this director, or of mixed character. Unlike the other cases, where the ground state is unique up to a global SU(2) rotation, the ground state
is highly degenerate in this case.

\begin{figure}[!t]
\includegraphics[width=7 truecm ]{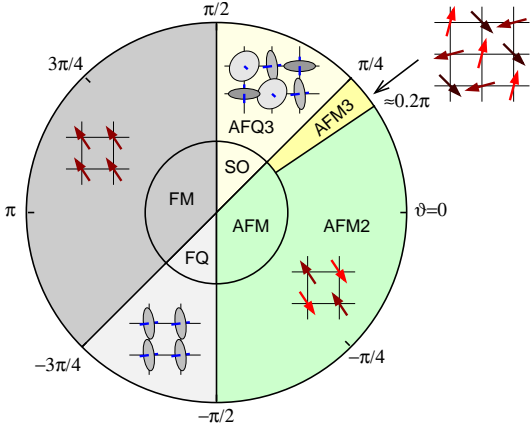}
\caption{(Color online) Schematic phase diagram of the $S=1$ bilinear-biquadratic model on the square lattice. The inner circle shows the variational, the outer the numerical results. FM and AFM are ferro- and antiferromagnetic, FQ is ferroquadrupolar, while SO stands for semiordered. In the outer circle, we have shown the three-sublattice ordered antiferromagnetic and antiferroquadrupolar phases between $\vartheta\approx 0.2 \pi$ and $\pi/2$.
\label{fig:phase_diag}}
\end{figure}

The above analysis leads to the zero-field classical phase diagram shown in Fig.~\ref{fig:phase_diag} (inner circle), with
the standard notation $J_1=J \cos \vartheta$ and $J_2=J \sin \vartheta$. It consists of
four phases, clockwise from the Heisenberg point: antiferromagnetic (AFM), ferroquadrupolar (FQ), ferromagnetic
(FM), and semiordered (SO) in the terminology of Papanicolaou.~\cite{papanicolaou88} The last phase is highly degenerate. For instance,
if all sites of one sublattice are in the same quadrupolar state, the sites of the other  sublattice can choose
their state (purely magnetic/quadrupolar or mixed) independently of each other. Another family
consists of all coverings of the square lattice using three quadrupoles with mutually perpendicular directors, with a residual
entropy equal to that of the 3-state Potts model on the square lattice.

\begin{figure}[!t]
\includegraphics[width=7 truecm]{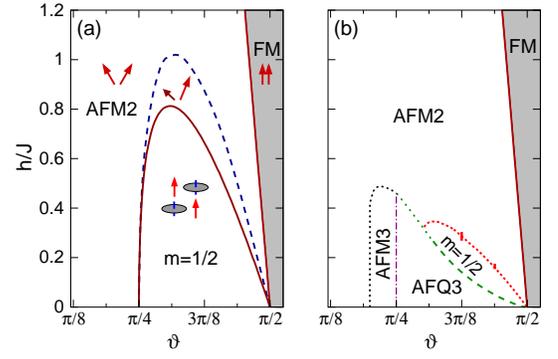}
\caption{(Color online) (a) The variational phase diagram in a magnetic field. AFM2 denotes a two-sublattice ordered N\'eel phase, while FM stands for ferromagnet. Note the presence of a magnetization plateau at 1/2 which is separated from AFM2 by a supersolid phase. (b) Schematic plot of the anticipated phase diagram based on exact diagonalization calculations. AFM3 (AFQ3) is a three-sublattice ordered antiferromagnetic (antiferroquadrupolar)  phase. The upper boundary of the 1/2 plateau is a result of finite-size scaling. The dashed line is estimated within flavor-wave theory (see text). \label{fig:PDh}}
\end{figure}

In a magnetic field, this degeneracy is lifted: The two-sublattice configuration with identical quadrupoles on
one sublattice and purely magnetic states with a moment along the field on the other sublattice is the
unique ground state. This gives rise to a magnetization plateau at 1/2 (see Fig.~\ref{fig:PDh}). Upon leaving the plateau,
the system adopts a two-sublattice configuration with mixed states: At first, these states feature different polarizations along
the field (a kind of spin supersolid), and subsequently a canted antiferromagnet is stabilized.

In summary, up to the degeneracy of the ground state in zero field for $\pi/4<\vartheta<\pi/2$, the classical
phase diagram consists of uniform or two-sublattice phases, with an abrupt transition into a 1/2 plateau for an infinitesimal field in this parameter range. As we shall now demonstrate, this picture is considerably affected
by quantum fluctuations.

\begin{figure}[b]
 \includegraphics[width=8truecm]{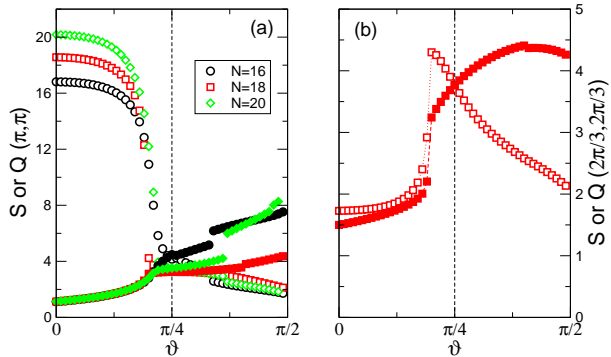}
\caption{(Color online) Spin and quadrupole structure factor from exact diagonalizations of square clusters with periodic boundary conditions. The geometry of the clusters can be found in Fig.~{\color{red} 72} of Ref.~\onlinecite{Dagotto1994}.
Open (closed) symbols stand for the spin (quadrupole) structure factor. \label{fig:SqQq}}
\end{figure}

\begin{figure}[h]
\includegraphics[width=8truecm]{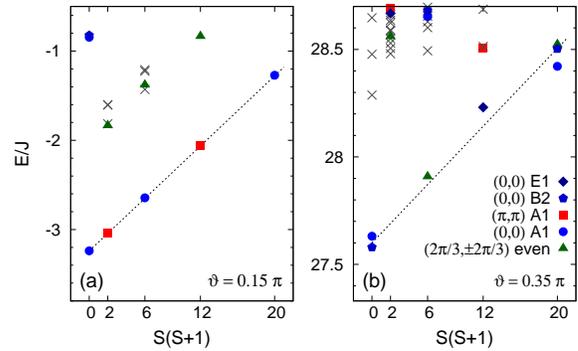}
\caption{(Color online) Anderson tower for the 18-site cluster showing the two-sublattice (a) and the three-sublattice order (b). The dashed line is a guide to the eye. The energy levels are characterized by their momenta and irreducible representations. \label{fig:AndersonTower}}
\end{figure}

{\it Numerical approach}. For $J_2<0$, there is no sign problem for quantum Monte Carlo simulations, and Harada and Kawashima~\cite{harada}
have shown that the sequence of phases (AFM, FQ and FM) as well as the location of the transitions predicted by the classical approach are preserved by
quantum fluctuations. In the following, we concentrate on the parameter range $J_1,J_2>0$ ($0<\vartheta<\pi/2$).
In order to get an insight into the nature of the zero-field phase, we have calculated the structure factors $\sum_j\exp[i\mathbf{k}\cdot\mathbf{R}_j]\langle C(0)\cdot C(\mathbf{R}_j)\rangle$, where $C(\mathbf{R}_j)$ is the spin or quadrupole operator~\cite{qpoperators} at site $\mathbf{R}_j$, by exact diagonalization for different wave vectors $\mathbf{k}$ as a function of $\vartheta$ (Fig.~\ref{fig:SqQq}). For $\vartheta =0$ (Heisenberg model), the structure factor is the largest (and grows with the system size) at $(\pi,\pi)$, as we expect for a two-sublattice ordered N\'eel antiferromagnet. As we turn on the biquadratic coupling, the structure factor gradually decreases, and for $\vartheta \agt 0.19 \pi$, i.e., long before the
classical transition point at $\pi/4$ is reached, three-sublattice correlations take over, with a structure factor that peaks at $(2\pi/3,2\pi/3)$. Spin-spin correlations prevail for $\vartheta<\pi/4$, while quadrupolar correlations become dominant in the region where $J_2>J_1$. A three-sublattice ordering is also suggested by the peculiar dependence of the ground-state energy on the number of sites: We consistently get lower energies for clusters that are multiples of three. Furthermore, for $\vartheta>\pi/4$, the low-energy spectrum of the 18-site cluster reveals an Anderson tower~\cite{AndersonPR52,BernuPRL92} (Fig.~\ref{fig:AndersonTower}) that corresponds to two copies of the Anderson tower of the three-sublattice antiferroquadrupolar state on the triangular lattice.~\cite{triangularT,triangularL,KarloAndreas} The two copies refer to the $Z_2$ degeneracy of the state, i.e., to the orientation of the stripes. We expect a finite-temperature ordering transition of Ising type to select the orientation of the stripes, while the antiferroquadrupolar order is only established at $T=0$.

\begin{figure}[h]
\includegraphics[width=8.5truecm]{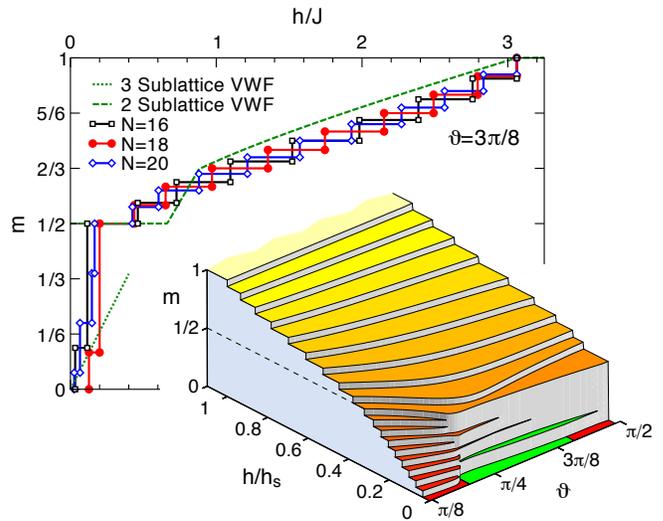}
\caption{(Color online) Magnetization curves of finite clusters for $\vartheta=3\pi/8$. Dashed lines represent variational magnetization curves calculated for two-sublattice and three-sublattice ordered states.
The 3D plot shows the magnetization process of the 18-site cluster as a function of $\vartheta$.
\label{fig:m3D}}
\end{figure}

We complete our numerical analysis by investigating the magnetization process in the range
$\pi/8<\vartheta<\pi/2$ (Fig.~\ref{fig:m3D}). In contrast to the classical solution, the magnetization does not jump directly to 1/2 for $\pi/4<\vartheta<\pi/2$: There is an intermediate phase at low field in which the system polarizes progressively and retains the three-sublattice order of the zero-field ground state. In order to see whether the plateau at 1/2 remains in the thermodynamic limit, we have attempted a finite-size scaling of its critical fields for 16, 18, and 20 sites. For the upper critical field, a linear finite-size scaling is possible for $\vartheta$ not too close to $\pi/4$, resulting
in the upper boundary of Fig.~\ref{fig:PDh}(b). For the lower critical field, however, no meaningful
finite-size scaling could be performed, presumably due to the incompatibility of the 16- and 20-site clusters with three-sublattice long-range order.

{\it Semiclassical approach}. The numerical results are confirmed and complemented by a semiclassical analysis that includes zero-point fluctuations
at the level of linear flavor-wave theory.~\cite{papanicolaou88,KarloAndreas,Joshi} In zero field for $\pi/4<\vartheta<\pi/2$, we have compared the energy of all two-sublattice ordered classical ground states and the three-sublattice antiferroquadrupolar state, and the latter state is indeed stabilized for all $\vartheta$. For $\vartheta=\pi/4$, this state is considerably lower in energy than the two-sublattice antiferromagnetic state, and even though we cannot follow the three-sublattice state below $\vartheta=\pi/4$, since it is no longer a classical ground state and imaginary frequencies appear in its flavor-wave spectrum, we may argue by continuity that three-sublattice order will be favored over two-sublattice antiferromagnetic
order in a finite window below $\vartheta=\pi/4$, in agreement with exact diagonalizations.
In a finite field for $\pi/4<\vartheta<\pi/2$, the 1/2-plateau state is the unique classical ground state and again we cannot calculate the energy of the three-sublattice state. However, for small fields, the correction to the energy is dominated by the classical energy gain of the 1/2-plateau state, which is linear in field and equal to $h/2$, since all
other corrections are quadratic. Therefore, the boundary between the three-sublattice phase and the 1/2 plateau is given by $h=2(\varepsilon_2- \varepsilon_3)$, where $\varepsilon_3$ (resp.~$\varepsilon_2$) is the zero-point correction to the energy of the three-sublattice antiferroquadrupolar (resp.~1/2-plateau) state calculated in zero field. This boundary is shown in Fig.~\ref{fig:PDh}(b). In the $\vartheta\rightarrow\pi/2$ limit, the critical field goes to zero and the approximation is quantitatively reliable.
Now, to leading order in $\delta = \pi/2-\vartheta$, the zero-point corrections are given by $\varepsilon_3 = -8 \delta^{3/2}/3\pi + O(\delta^2)$ and $\varepsilon_2=\delta^2 \ln \delta$; therefore the critical field scales as $h \propto \delta^{3/2}$. This implies that the lower boundary of the plateau has a vanishing slope in the $\vartheta\rightarrow\pi/2$ limit. Since the finite-size scaling of the upper critical
field deduced from exact diagonalizations is very good in the same limit and yields a significant slope, one can
safely conclude that a 1/2 plateau is indeed present in the phase diagram.
However, the fate of this plateau upon approaching $\vartheta=\pi/4$ cannot be deduced from this argument,
and the corresponding region of the phase diagram of Fig.~\ref{fig:PDh} is just a plausible scenario.
Note that both the semiclassical calculation and exact diagonalizations are consistent with
a magnetization jump at the lower critical field of the 1/2 plateau.

\begin{figure}[!t]
\includegraphics[width=8.5truecm]{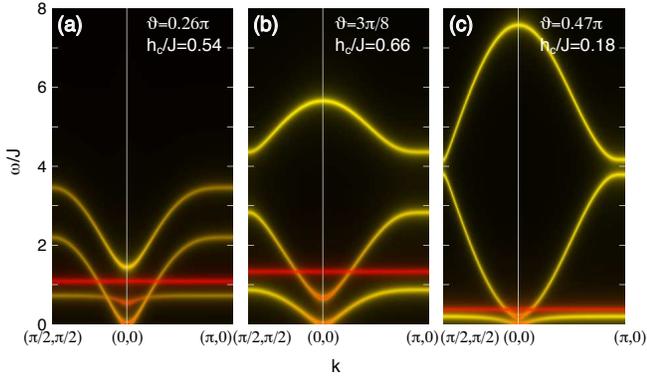}
\caption{(Color online)
 Dynamical structure factors for the plateau state in the reduced Brillouin zone for three different values of $\vartheta$ at the upper critical magnetic field where the gap closes. The color code goes from yellow for spin
 waves to red for quadrupolar waves according to the dominant character of the mode. The dispersing bands have finite $S^x=S^y$ and $Q^{yz}=Q^{zx}$ matrix elements. The flat dispersion at $\omega=2 h$ is a pure quadrupolar wave with $Q^{xy}=Q^{x^2-y^2}$ fluctuations. The spectra are displayed at the critical field to show that it
is a partially magnetic mode that softens at the transition.
\label{fig:struc_factors_m1}}
\end{figure}

Finally, let us comment on the nature of the elementary excitations. It is best revealed by the dynamical structure factors
\begin{equation}
C(\mathbf{k},\omega)=\sum_{j,\nu} |\langle\nu|C_j(\mathbf{k})|G\rangle|^2\delta(\omega-\omega_\nu),
\end{equation}
where $C_j(\mathbf{k})$ can be the Fourier transform of a spin or a quadrupole operator on sublattice $j$, the ground state is denoted by $|G\rangle$, and $|\nu\rangle$ are excited states with energy $\omega_\nu$.
In general, the excitations are of mixed character; i.e., they appear with measurable weight both in the spin
and quadrupole dynamical structure factors, with on average a stronger spin resp.~quadrupolar character
depending on the type of order, on the band, and on the wave vector.
For the 1/2-plateau phase, however, the flavor-wave analysis predicts that among the four bands, one
is dispersionless and of {\it pure quadrupolar} character for all $\vartheta$, as shown in Fig.~\ref{fig:struc_factors_m1}. This mode would be completely absent in neutron scattering, but could
be detected, e.g., in Raman scattering.~\cite{michaud}

The interpretation of the spectrum is particularly transparent for $\vartheta=\pi/2$, where the variational plateau state becomes an exact ground state. Restricting ourselves to single-particle states, we find that the $|z\rangle \rightarrow |1\rangle$ (with $\Delta S^z=+1$) and $|1\rangle \rightarrow |\bar 1\rangle$ ($\Delta S^z=-2$) excitations are localized, forming flat bands. The $|z\rangle \rightarrow |\bar1\rangle$ and $|1\rangle \rightarrow |z\rangle$ excitations ($\Delta S^z=-1$) on the two sublattices hybridize and form two bands with dispersion
$[4\pm2(\cos k_x + \cos k_y)]J$.
These excitations are recovered in the flavor-wave calculations in the $\vartheta\rightarrow\pi/2$ limit, and can be recognized in Fig.~\ref{fig:struc_factors_m1}(c). Away from $\vartheta=\pi/2$, the
$|z\rangle \rightarrow |1\rangle$ excitation acquires a dispersion, while the $|1\rangle \rightarrow |\bar 1\rangle$ quadrupolar excitation remains dispersionless.

{\it Conclusions}.
 In summary, the evidence presented here strongly suggests that the bilinear-biquadratic Heisenberg model on the
square lattice features a phase with three-sublattice order, although
the lattice is bipartite and the coupling is limited to nearest neighbors. The extent of this phase is
quite surprising in view of the classical phase diagram: It replaces a large portion of the 1/2 plateau in a
field, and it extends to the region where the exchange is predominantly bilinear. The 1/2 plateau is also
quite interesting, with a purely quadrupolar dispersionless excitation. It is our hope that the present
work will further motivate the search for $S=1$ quantum magnets with significant positive biquadratic
interactions.

\acknowledgements
The authors are grateful to S.~Manmana for discussions. This work was supported by the Swiss National Fund, by MaNEP, and by Hungarian OTKA Grant No. K73455.


\begin{thebibliography}{99}
\bibitem{qpoperators} The quadrupole operator has five components: $Q^{x^2-y^2}=(S^x)^2 - (S^y)^2$, $Q^{3z^2-r^2}=\left[2 (S^z)^2 - (S^x)^2 - (S^y)^2\right]/\sqrt{3}$, $Q^{xy}=S^x S^y + S^y S^x$, $Q^{yz}=S^y S^z + S^z S^y$, and $Q^{zx}=S^z S^x + S^x S^z$.
\end{thebibliography}
\end{document}